\documentclass{jkas}

\def\year{2015} 
\def\month{October} 
\def\volume{48} 
\def\issue{5} 
\def\beginpage{267} 
\def\endpage{275} 
\def\received{September 21, 2015} 
\def\accepted{October 13, 2015} 
\date{Received \received; accepted \accepted}

\newcommand{\hh}{\mbox{\hspace{1.5mm}}}

\newcommand{\mm}{\mbox{\hspace{2.5mm}}}
\newcommand{\x}{\mbox{\hspace{3.0mm}}}

\newcommand{\e}{\mbox{\hspace{4.0mm}}}

\usepackage[usenames]{color}

\title{KVN Monitoring Observations toward the Recent Outburst Symbiotic Star V407 Cygni}

\author[]{Se-Hyung Cho}
\author[]{Jaeheon Kim}
\author[]{Youngjoo Yun}

\affil[]{Radio Astronomy Division, Korea Astronomy and Space Science Institute, 776 Daedukdae-ro, Yuseong-gu, Daejeon 305-348, Republic of Korea; 
\email{cho@kasi.re.kr, jhkim@kasi.re.kr, yjyun@kasi.re.kr}}

\def\corrauthor{S.-H. Cho}

\def\runningauthor{Cho, Kim, \& Yun}

\def\runningtitle{KVN Monitoring Observation of V407 Cyg}

\def\keywords{circumstellar matter: masers: radio lines: stars --- symbiotic star}

\def\abstracttext{
Simultaneous time monitoring observations of H$_{2}$O and SiO maser lines were performed toward the D-type symbiotic binary system V407 Cyg 
with the Korean VLBI Network single dish radio telescope. 
These monitoring observations were carried out from March 2, 2010 (optical phase $\phi$ = 0.0), 8 days before the nova outburst on March 10, 2010 to June 5, 2014 ($\phi$ = 2.13). 
Eight days before the nova outburst, we detected the SiO $v$ = 1, 2, $J$ = 1--0 maser lines which exhibited values of 0.51 K ($\sim$ 6.70 Jy) and 0.71 K ($\sim$ 9.30 Jy), 
respectively, while after the outburst we could not detect them on April 2 ($\phi$ = 0.04), May 5 ($\phi$ = 0.09), May 8 ($\phi$ = 0.09), or on June 5, 2010  ($\phi$ = 0.13) 
within the upper limits of our KVN observations. 
After restarting our monitoring observations, we detected SiO $v$ = 2, $J$ = 1--0 masers starting on October 20, 2011 ($\phi$ = 0.83) 
and detected SiO $v$ = 1, $J$ = 1--0 masers starting on December 22, 2011 ($\phi$ = 0.92). 
These results provide clear evidence of the interaction between the shock from the nova outburst and the SiO maser regions of the Mira envelope. 
The peak emission of SiO $v$ = 1, 2, $J$ = 1--0 masers always occurred at blueshifted velocities with respect to the stellar velocity except for that of SiO $v$ = 1 at one epoch. 
These phenomena may be related to the redistribution of SiO maser regions after the outburst. 
The peak velocity variations of SiO masers associated with stellar pulsation phases show an increasing blueshifted trend during our monitoring interval after the outburst. 
}

\begin{document}
\jkashead 

\section{Introduction\label{sec:intro}}

\indent\indent V407 Cyg is a D-type (dust-rich) symbiotic binary system consisting of a hot companion white dwarf and a cool companion M6 III Mira-type variable star 
with a 745 day pulsation period \citep{mun90}. The distance is estimated to be 2.7 kpc from the Mira period-luminosity relation, and the corresponding binary separation 
between a white dwarf and a Mira-type star is 17 AU with a 43 year orbital period in the long-term optical light curve \citep{mun90}. 
The hot compact companion is accreting matter from the Mira-type star through the stellar wind. 
These accreting matters cause nova outbursts in the hydrogen-rich envelope of a white dwarf as a thermonuclear runaway explosion. 
The outburst of V407 Cyg was discovered on March 10, 2010 by \citet{mae10} at V = 7.6 mag, showing an unsurpassed brightness level in the photometric history of the system. 
After the outburst, follow-up observations were performed in several wavelength regimes, for example radio continuum \citep{bow10,gir10,chom12}, SiO maser \citep{deg11}, 
optical \citep{mun11,sho11,iij15}, infrared \citep{jos10}, X-rays and gamma rays \citep{abd10,nel12}. 
In particular, the gamma-ray detection from V407 was the first reported detection in this wavelength from any nova; this was explained as particle accelerations caused by interaction of the nova shock 
with the dense Mira wind \citep{abd10}. \citet{mun11}, based on both photometric and spectrometric observations in optical and infrared wavelength regimes, reported 
that the nova ejecta is progressively decelerating with time as it is expanding into the Mira wind from a FWHM of 2760 km s$^{-1}$ (+2.3 days after the outburst) 
to 200 km s$^{-1}$ (+196 days). \citet{deg11} found the time variation of SiO maser line profiles based on monitoring observations of 2.5 months after the outburst. 
They indicated that the SiO emitting regions were wiped out by the nova shock, 
but a part of the maser regions was quickly replenished by cool molecular gases expelled by the pulsation of the Mira. 
Using the KVN single dish radio telescope, we have performed time monitoring observations toward the symbiotic star V407 Cyg starting on March 2, 2010 before the nova outburst. 
Here, we present the results of simultaneous monitoring observations of H$_{2}$O and SiO masers toward V407 Cyg. 
In Section 2, we present the observations. Observational results and a discussion are given in Sections 3 and 4, respectively. A summary is given in Section 5. 

\section{Observations\label{sec:obs}}

\indent\indent We started simultaneous monitoring observations of H$_{2}$O and SiO masers toward V407 Cyg from March 2, 2010 (optical phase $\phi$ = 0.0), 
8 days before the outburst. The monitoring  observations were performed every month until June 5, 2010 ($\phi$ = 0.13). 
However monitoring was stopped during the maintenance season of July-August 2010 to September 2011, after which it restarted on October 20, 2011 ($\phi$ = 0.83), operating until June 5, 2014 ($\phi$ = 2.13). 
Observations of H$_{2}$O $6_{16}-5_{23}$ and SiO $J$ = 1--0 maser lines were made from the first stage of KVN single dish operation starting in 2009. 
Monitoring observations of SiO $J$ = 2--1, $J$ = 3--2 lines were added starting in May 2012. 
The thermal lines of $^{28}$SiO $v$ = 0, $J$ = 1--0, $J$ = 2--1 and high vibrational maser lines of $v$ = 3, 4, $J$ = 1--0 were searched together with silicon isotopic lines of 
$^{29}$SiO $v$ = 0, 1, $J$ = 1--0 and $^{30}$SiO $v$ = 0, $J$ = 1--0. 
The H$_{2}$O and SiO transitions and rest frequencies used for these observations are listed in Table~\ref{tab:jkastable1}. 


\begin{table}[tp]
\caption{H$_{2}$O and SiO transitions and rest frequencies used for the observations\label{tab:jkastable1}}
\centering
\begin{tabular}{ccr}

\toprule
Molecule      & Transition       & Frequency (GHz) \\

\midrule
H$_{2}$O      & 6$_{1,6}$--5$_{2,3}$ &  22.235080\e \\
$^{28}$SiO\hh & $v$=0, $J$=1--0      &  43.423858\e \\
              & $v$=1, $J$=1--0      &  43.122080\e \\
              & $v$=2, $J$=1--0      &  42.820587\e \\
              & $v$=3, $J$=1--0      &  42.519379\e \\
              & $v$=4, $J$=1--0      &  42.218456\e \\
              & $v$=0, $J$=2--1      &  86.846998\e \\
              & $v$=1, $J$=2--1      &  86.243442\e \\
              & $v$=2, $J$=2--1      &  85.640452\e \\
              & $v$=1, $J$=3--2      & 129.363359\e \\
              & $v$=2, $J$=3--2      & 128.458891\e \\
$^{29}$SiO\hh & $v$=0, $J$=1--0      &  42.879916\e \\
              & $v$=1, $J$=1--0      &  42.583798\e \\
$^{30}$SiO\hh & $v$=0, $J$=1--0      &  42.373359\e \\

\bottomrule

\end{tabular}
\end{table}


One from among the KVN Yonsei, Ulsan, and Tamna 21 m single dish telescopes was used according to the observational dates. 
The KVN antenna optics were designed to support simultaneous observations at four wavelength regimes at the 22, 43, 86, and 129 GHz bands 
using three quasi-optical low-pass filters \citep{han08,han13}. 
The half power beam widths and aperture efficiencies of the three KVN telescopes at the four wavelength bands are given on the KVN home page (\url{http://kvn-web.kasi.re.kr/}). 
The average values of the three telescopes were adopted because their antenna parameters are similar. 
Hence, the average half power beam widths and aperture efficiencies are 123$''$, 0.58 (at 22 GHz), 62$''$, 0.61 (at 43 GHz), 32$''$, 0.50 (at 86 GHz) 
and 23$''$, 0.35 (at 129 GHz), respectively. The pointing accuracy was checked every two hours using a strong SiO maser source, $\chi$ Cyg. 
The cryogenic 22, 43, and 86 GHz High Electron Mobility Transistor (HEMT) receivers and the Superconductor-Insulator-Superconductor (SIS) 129 GHz receiver 
which can be operated with both right and left circular polarized feeds \citep{han13} were used. 
We used only the left circular polarized feed during our observations. 

The system noise temperatures of a single side-band range from 83 K to 241 K (at 22 GHz), from 132 K to 276 K (at 43 GHz), from 196 K to 395 K (at 86 GHz), 
and from 279 K to 573 K (at 129 GHz) depending on weather conditions and elevations. 
As a standard setting for our monitoring projects, such as the monitoring of TX Cam \citep{cho14}, we used a digital spectrometer with the total band widths chosen 
from one 64 MHz mode for the H$_{2}$O $6_{16}-5_{23}$ line and three 64 MHz modes for the SiO $v$ = 1, 2, and $^{29}$SiO $v$ = 0, $J$ = 1--0 lines 
until the April 2012 observations. 
After that period, for the four bands (22, 43, 86, and 129 GHz), we used three 32 MHz modes for the H$_{2}$O $6_{16}-5_{23}$, SiO $v$ = 1, 2, $J$ = 1--0 lines 
and two 64 MHz modes for the SiO $v$ = 1, $J$ = 2--1 and $J$ = 3--2 lines. 
These band widths correspond to radial velocity ranges of 860 km s$^{-1}$ (at 22 GHz) and 440 km s$^{-1}$ (at 43 GHz). 
The velocity resolutions of each band are 0.21 km s$^{-1}$ (at 22 GHz) and 0.11 km s$^{-1}$ (at 43 GHz) until the April 2012 observation mode. 
After that, for the four band modes, the radial velocity ranges are 440 km s$^{-1}$ (at 22 GHz), 222 km s$^{-1}$ (at 43 GHz), 222 km s$^{-1}$ (at 86 GHz), 
and 148 km s$^{-1}$ (at 129 GHz) with velocity resolutions of 0.11 km s$^{-1}$ (at 22 GHz), 0.05 km s$^{-1}$ (at 43 GHz), 0.05 km s$^{-1}$ (at 86 GHz), 
and 0.036 km s$^{-1}$ (at 129 GHz), respectively. All spectra were Hanning-smoothed to velocity resolutions of 0.44--0.57 km s$^{-1}$.

We used the chopper wheel method for data calibration; this method corrects the atmospheric attenuation and the antenna gain variations, 
depending on the elevation in order to yield the antenna temperature $T^\ast_A$. 
The integration time was 60$-$120 minutes to achieve the sensitivity of $\sim$0.03--0.09 K at the 3$\sigma$ level. 
The average conversion factors of the three telescopes from the antenna temperature to the flux density were approximately 13.8 Jy K$^{-1}$ at 22 GHz, 13.1 Jy K$^{-1}$ at 43 GHz, 
15.9 Jy K$^{-1}$ at 86 GHz, and 22.8 Jy K$^{-1}$ at 129 GHz, respectively.

\section{Observational Results\label{sec:result}}


\begin{figure*}[tp]
\centering
\includegraphics[width=160mm]{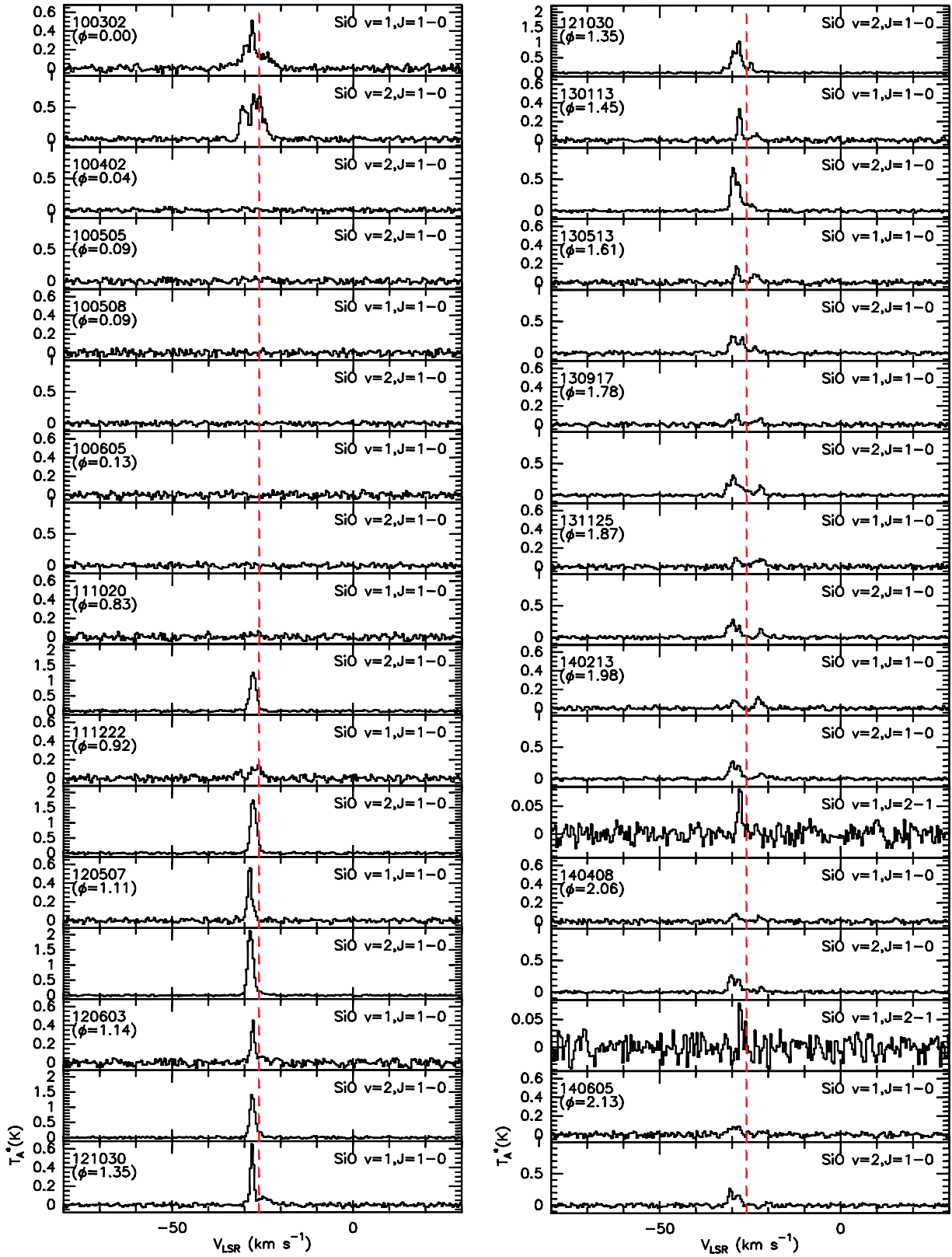}
\caption{Simultaneously obtained spectra of the SiO $v$=1, 2, $J$=1--0 and $v$=1, $J$=2--1 masers from March 02, 2010 to June 05, 2014 (17 epochs). 
The intensity is given in units of the antenna temperature $T^\ast_A$ (K) and the abscissa is $V_{\rm LSR}$ (km s$^{-1}$). 
The date of observation (yymmdd), the optical phase, and the transition of molecule are also given for each spectrum. \label{fig:jkasfig1}}
\end{figure*}


\indent\indent Figure~\ref{fig:jkasfig1} displays the spectra of the monitoring observations centered on the SiO $v$ = 1, 2, $J$ = 1--0 masers from March 2, 2010 
before the nova outburst (March 10, on 2010) to June 5, 2014 after the nova outburst. 
The SiO maser spectra on March 2, 2010 ($\phi$ = 0.0), only 8 days before the outburst exhibited 
0.51 K ($\sim$ 6.70 Jy) in the $v$ = 1, $J$ = 1--0 and 0.71 K ($\sim$ 9.30 Jy) in the $v$ = 2, $J$ = 1--0 lines, 
while those spectra were not detected on April 2 ($\phi$ = 0.04), May 5 ($\phi$ = 0.09), May 8 ($\phi$ = 0.09), and June 5, 2010 ($\phi$ = 0.13) 
within the upper limits of $\sim$ 0.78--1.18 Jy after the burst. After the discontinued period of monitoring from July of 2010 to September of 2011, 
the SiO $v$ = 2, $J$ = 1--0 masers began to be detected starting on October 20, 2011 ($\phi$ = 0.83). 
The SiO $v$ = 1, $J$ = 1--0 masers also began to be detected on December 22, 2011 ($\phi$ = 0.92). 
The spectra three months before the outburst (December 2, 2009) were also obtained by \citet{cho10}. 


\begin{table*}
\caption{Results of the H$_{2}$O and SiO Maser Monitoring Observations\label{tab:jkastable2}}
\centering
\begin{tabular}{ccccccc}

\toprule
Molecular  & $T_{\rm A}^{\ast}$(peak) & rms & \x $\int\! T_{\rm A}^{\ast}dv$\e & $V_{\rm LSR}$(peak) & $V_{\rm LSR}$(peak)$-V^{\ast}$ & Date(phase) \\
Transition & (K)                      & (K) & (K km s$^{-1}$)                  & (km s$^{-1}$)       & (km s$^{-1}$)                  & (yymmdd)    \\
(1)        & (2)                      & (3) & (4)                              & (5)                 & (6)                            & (7)         \\

\midrule							
H$_{2}$O 6$_{1,6}$--5$_{2,3}$ & $\cdots$ & 0.02 & $\cdots$ & $\cdots$ & $\cdots$ & 100302(0.00) \\
                              & $\cdots$ & 0.02 & $\cdots$ & $\cdots$ & $\cdots$ & 100402(0.04) \\
                              & $\cdots$ & 0.03 & $\cdots$ & $\cdots$ & $\cdots$ & 100505(0.09) \\
                              & $\cdots$ & 0.03 & $\cdots$ & $\cdots$ & $\cdots$ & 100508(0.09) \\
                              & $\cdots$ & 0.03 & $\cdots$ & $\cdots$ & $\cdots$ & 100516(0.10) \\
                              & $\cdots$ & 0.03 & $\cdots$ & $\cdots$ & $\cdots$ & 100605(0.13) \\
                              & $\cdots$ & 0.02 & $\cdots$ & $\cdots$ & $\cdots$ & 111020(0.83) \\
                              & $\cdots$ & 0.02 & $\cdots$ & $\cdots$ & $\cdots$ & 111222(0.92) \\
                              & $\cdots$ & 0.02 & $\cdots$ & $\cdots$ & $\cdots$ & 120507(1.11) \\
                              & $\cdots$ & 0.03 & $\cdots$ & $\cdots$ & $\cdots$ & 120603(1.14) \\
                              & $\cdots$ & 0.01 & $\cdots$ & $\cdots$ & $\cdots$ & 121030(1.35) \\
                              & $\cdots$ & 0.01 & $\cdots$ & $\cdots$ & $\cdots$ & 130113(1.45) \\
                              & $\cdots$ & 0.02 & $\cdots$ & $\cdots$ & $\cdots$ & 130513(1.61) \\
                              & $\cdots$ & 0.02 & $\cdots$ & $\cdots$ & $\cdots$ & 130917(1.78) \\
                              & $\cdots$ & 0.02 & $\cdots$ & $\cdots$ & $\cdots$ & 131125(1.87) \\
                              & $\cdots$ & 0.01 & $\cdots$ & $\cdots$ & $\cdots$ & 140213(1.98) \\
                              & $\cdots$ & 0.02 & $\cdots$ & $\cdots$ & $\cdots$ & 140408(2.06) \\
                              & $\cdots$ & 0.03 & $\cdots$ & $\cdots$ & $\cdots$ & 140605(2.13) \\
\addlinespace                             
$^{28}$SiO $v$=0,$J$=1--0     & $\cdots$ & 0.02 & $\cdots$ & $\cdots$ & $\cdots$ & 120507(1.11) \\
                              & $\cdots$ & 0.03 & $\cdots$ & $\cdots$ & $\cdots$ & 120603(1.14) \\
\addlinespace                          
$^{28}$SiO $v$=1,$J$=1--0     & 0.51     & 0.02 & 2.01     & $-$27.9  & $-$1.9   & 100302(0.00) \\
                              & $\cdots$ & 0.02 & $\cdots$ & $\cdots$ & $\cdots$ & 100508(0.09) \\
                              & $\cdots$ & 0.02 & $\cdots$ & $\cdots$ & $\cdots$ & 100605(0.13) \\
                              & $\cdots$ & 0.02 & $\cdots$ & $\cdots$ & $\cdots$ & 111020(0.83) \\
                              & 0.14     & 0.02 & 0.55     & $-$26.3  & $-$0.3   & 111222(0.92) \\
                              & 0.56     & 0.02 & 0.90     & $-$27.9  & $-$1.9   & 120507(1.11) \\
                              & 0.46     & 0.03 & 0.82     & $-$27.6  & $-$1.6   & 120603(1.14) \\
                              & 0.65     & 0.01 & 1.14     & $-$27.9  & $-$1.9   & 121030(1.35) \\
                              & 0.34     & 0.02 & 0.58     & $-$27.9  & $-$1.9   & 130113(1.45) \\
                              & 0.17     & 0.02 & 0.45     & $-$28.8  & $-$2.8   & 130513(1.61) \\
                              & 0.12     & 0.01 & 0.37     & $-$28.4  & $-$2.4   & 130917(1.78) \\
                              & 0.10     & 0.02 & 0.47     & $-$28.8  & $-$2.8   & 131125(1.87) \\
                              & 0.12     & 0.01 & 0.42     & $-$22.8  & \mm 3.2  & 140213(1.98) \\
                              & 0.08     & 0.02 & 0.33     & $-$28.8  & $-$2.8   & 140408(2.06) \\
                              & 0.09     & 0.02 & 0.41     & $-$27.9  & $-$1.9   & 140605(2.13) \\
\addlinespace                          
$^{28}$SiO $v$=2,$J$=1--0     & 0.71     & 0.02 & 3.66     & $-$27.5  & $-$1.5   & 100302(0.00) \\
                              & $\cdots$ & 0.02 & $\cdots$ & $\cdots$ & $\cdots$ & 100402(0.04) \\
                              & $\cdots$ & 0.03 & $\cdots$ & $\cdots$ & $\cdots$ & 100505(0.09) \\
                              & $\cdots$ & 0.02 & $\cdots$ & $\cdots$ & $\cdots$ & 100508(0.09) \\
                              & $\cdots$ & 0.03 & $\cdots$ & $\cdots$ & $\cdots$ & 100605(0.13) \\
                              & 1.27     & 0.02 & 2.81     & $-$27.6  & $-$1.6   & 111020(0.83) \\
                              & 1.74     & 0.02 & 3.67     & $-$27.6  & $-$1.6   & 111222(0.92) \\
                              & 2.13     & 0.02 & 4.35     & $-$27.9  & $-$1.9   & 120507(1.11) \\
                              & 1.41     & 0.02 & 2.72     & $-$28.0  & $-$2.0   & 120603(1.14) \\
                              & 1.04     & 0.01 & 3.67     & $-$28.0  & $-$2.0   & 121030(1.35) \\
                              & 0.68     & 0.01 & 2.02     & $-$29.7  & $-$3.7   & 130113(1.45) \\
                              & 0.27     & 0.02 & 1.19     & $-$30.1  & $-$4.1   & 130513(1.61) \\
                              & 0.31     & 0.01 & 1.38     & $-$29.7  & $-$3.7   & 130917(1.78) \\
                              & 0.28     & 0.02 & 1.09     & $-$29.8  & $-$3.8   & 131125(1.87) \\
                              & 0.28     & 0.01 & 1.09     & $-$29.8  & $-$3.8   & 140213(1.98) \\
                              & 0.27     & 0.01 & 0.99     & $-$30.1  & $-$4.1   & 140408(2.06) \\
                              & 0.27     & 0.02 & 0.67     & $-$30.6  & $-$4.6   & 140605(2.13) \\
\addlinespace                          
$^{28}$SiO $v$=3,$J$=1--0     & $\cdots$ & 0.02 & $\cdots$ & $\cdots$ & $\cdots$ & 100402(0.04) \\
                              & $\cdots$ & 0.03 & $\cdots$ & $\cdots$ & $\cdots$ & 100505(0.09) \\
                              & $\cdots$ & 0.03 & $\cdots$ & $\cdots$ & $\cdots$ & 100605(0.13) \\

\bottomrule               
                          
\end{tabular}             
\end{table*}              
                          
\setcounter{table}{1}     
\begin{table*}[t!]
\caption{(Continued)\label{tab:jkastable2-cont}}
\centering
\begin{tabular}{ccccccc}

\toprule
Molecular  & $T_{\rm A}^{\ast}$(peak) & rms & \x $\int\! T_{\rm A}^{\ast}dv$\e & $V_{\rm LSR}$(peak) & $V_{\rm LSR}$(peak)$-V^{\ast}$ & Date(phase) \\
Transition & (K)                      & (K) & (K km s$^{-1}$)                  & (km s$^{-1}$)       & (km s$^{-1}$)                  & (yymmdd)    \\
(1)        & (2)                      & (3) & (4)                              & (5)                 & (6)                            & (7)         \\

\midrule							                       
$^{28}$SiO $v$=4,$J$=1--0 & $\cdots$ & 0.03 & $\cdots$ & $\cdots$ & $\cdots$ & 100516(0.10) \\
                          & $\cdots$ & 0.03 & $\cdots$ & $\cdots$ & $\cdots$ & 100605(0.13) \\
\addlinespace
$^{28}$SiO $v$=0,$J$=2--1 & $\cdots$ & 0.02 & $\cdots$ & $\cdots$ & $\cdots$ & 120507(1.11) \\
\addlinespace
$^{28}$SiO $v$=1,$J$=2--1 & $\cdots$ & 0.02 & $\cdots$ & $\cdots$ & $\cdots$ & 120507(1.11) \\
                          & $\cdots$ & 0.02 & $\cdots$ & $\cdots$ & $\cdots$ & 120603(1.14) \\
                          & $\cdots$ & 0.01 & $\cdots$ & $\cdots$ & $\cdots$ & 121030(1.35) \\
                          & $\cdots$ & 0.01 & $\cdots$ & $\cdots$ & $\cdots$ & 130113(1.45) \\
                          & $\cdots$ & 0.02 & $\cdots$ & $\cdots$ & $\cdots$ & 130513(1.61) \\
                          & $\cdots$ & 0.02 & $\cdots$ & $\cdots$ & $\cdots$ & 130917(1.78) \\
                          & $\cdots$ & 0.02 & $\cdots$ & $\cdots$ & $\cdots$ & 131125(1.87) \\
                          & 0.08     & 0.01 & 0.12     & $-$27.9  & $-$1.9   & 140213(1.98) \\
                          & 0.08     & 0.02 & 0.04     & $-$27.9  & $-$1.9   & 140408(2.06) \\
                          & $\cdots$ & 0.03 & $\cdots$ & $\cdots$ & $\cdots$ & 140605(2.13) \\
\addlinespace                         
$^{28}$SiO $v$=2,$J$=2--1 & $\cdots$ & 0.03 & $\cdots$ & $\cdots$ & $\cdots$ & 120603(1.14) \\
\addlinespace
$^{28}$SiO $v$=1,$J$=3--2 & $\cdots$ & 0.04 & $\cdots$ & $\cdots$ & $\cdots$ & 120603(1.14) \\
                          & $\cdots$ & 0.03 & $\cdots$ & $\cdots$ & $\cdots$ & 130513(1.61) \\
                          & $\cdots$ & 0.02 & $\cdots$ & $\cdots$ & $\cdots$ & 130917(1.78) \\
                          & $\cdots$ & 0.02 & $\cdots$ & $\cdots$ & $\cdots$ & 131125(1.87) \\
                          & $\cdots$ & 0.02 & $\cdots$ & $\cdots$ & $\cdots$ & 140213(1.98) \\
                          & $\cdots$ & 0.01 & $\cdots$ & $\cdots$ & $\cdots$ & 140408(2.06) \\
                          & $\cdots$ & 0.04 & $\cdots$ & $\cdots$ & $\cdots$ & 140605(2.13) \\
\addlinespace                          
$^{28}$SiO $v$=2,$J$=3--2 & $\cdots$ & 0.05 & $\cdots$ & $\cdots$ & $\cdots$ & 120603(1.14) \\
\addlinespace
$^{29}$SiO $v$=0,$J$=1--0 & $\cdots$ & 0.02 & $\cdots$ & $\cdots$ & $\cdots$ & 100302(0.00) \\
                          & $\cdots$ & 0.03 & $\cdots$ & $\cdots$ & $\cdots$ & 100402(0.04) \\
                          & $\cdots$ & 0.03 & $\cdots$ & $\cdots$ & $\cdots$ & 100505(0.09) \\
                          & $\cdots$ & 0.02 & $\cdots$ & $\cdots$ & $\cdots$ & 100508(0.09) \\
                          & $\cdots$ & 0.03 & $\cdots$ & $\cdots$ & $\cdots$ & 100605(0.13) \\
                          & $\cdots$ & 0.02 & $\cdots$ & $\cdots$ & $\cdots$ & 111020(0.83) \\
                          & $\cdots$ & 0.02 & $\cdots$ & $\cdots$ & $\cdots$ & 111222(0.92) \\
\addlinespace
$^{29}$SiO $v$=1,$J$=1--0 & $\cdots$ & 0.03 & $\cdots$ & $\cdots$ & $\cdots$ & 100516(0.10) \\
                          & $\cdots$ & 0.03 & $\cdots$ & $\cdots$ & $\cdots$ & 100605(0.13) \\
\addlinespace                          
$^{30}$SiO $v$=0,$J$=1--0 & $\cdots$ & 0.03 & $\cdots$ & $\cdots$ & $\cdots$ & 100516(0.10) \\
                          & $\cdots$ & 0.03 & $\cdots$ & $\cdots$ & $\cdots$ & 100605(0.13) \\
                          
\bottomrule

\end{tabular}
\end{table*}


These results provide clear evidence of the interaction of the shock from the nova outburst with the envelope of Mira as shown by \citet{abd10} and \citet{deg11}. 
\citet{deg11}, using the Nobeyama Radio Observatory (NRO) 45 m telescope, detected the weaker emission of the $v$ = 1 (0.44 Jy) and $v$ = 2 (1.0 Jy) masers on March 16, 2010 
after the outburst as compared to the previous emission observed on April 18, 2002 (6.32 Jy in the $v$ = 1 line, 3.36 Jy in the $v$ = 2 line, \citealt{deg05}). 

These weak intensities were not detected from our KVN monitoring observations within three rms detection levels of $\sim$ 0.78--1.18 Jy (upper limits) during this period. 
\citet{deg11} interpreted these variations of the SiO line profiles and intensities using a shock propagation model of the nova outburst. 
The SiO $v$ = 1, $J$ = 2--1 maser was also detected on February 13, 2014 ($\phi$ = 1.98) and on April 8, 2014 ($\phi$ = 2.06) for the first time 
showing a blueshifted emission ($V_{\rm LSR}$ = $-$27.9 km s$^{-1}$) with respect to the stellar velocity ($V_{\rm LSR}$ = $-$26 km s$^{-1}$). 


\begin{figure*}[t!]
\centering
\includegraphics[width=140mm]{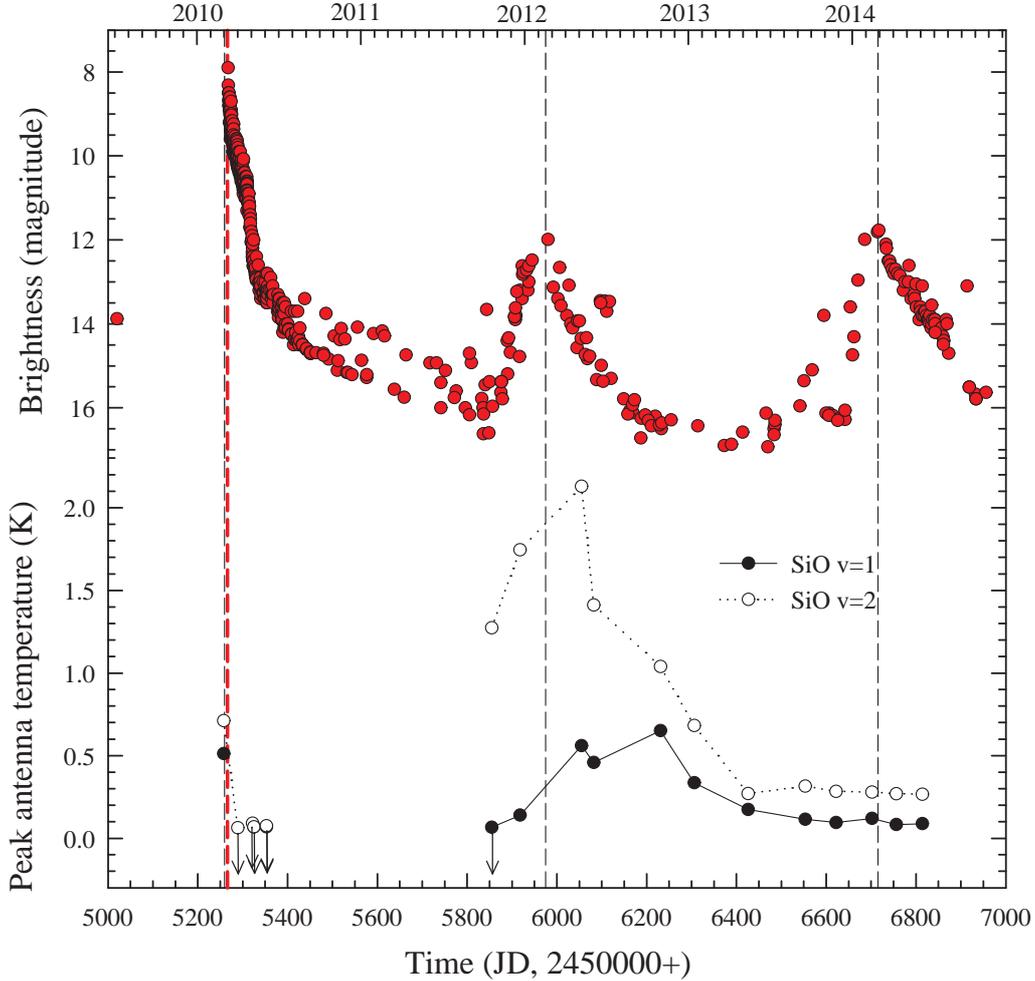}
\caption{The peak antenna temperature variations of the SiO $v$ = 1 and $v$ = 2, $J$ = 1--0 masers (black and open circles) are plotted on the optical light curve of V407 Cyg 
from AAVSO ({\color {red} red circles}) during our monitoring observations from March 2, 2010 to June 5, 2014. Optical maxima are indicated by the dashed vertical lines shown in black. 
The arrows indicate the upper limits of the SiO masers. The dashed vertical line shown in {\color {red} red} indicates the nova outburst epoch of March 10, 2010. 
The X-axis denotes Julian dates. \label{fig:jkasfig2}}
\end{figure*}


Table~\ref{tab:jkastable2} presents all of the observed properties of H$_{2}$O and SiO monitoring observations. 
The observed molecules and transitions are given in Column 1. The peak antenna temperatures and their root mean square (rms) levels are given in Columns 2 and 3, respectively. 
Columns 4 and 5 provide the integrated antenna temperatures and the peak $V_{\rm LSR}$ velocities, respectively. 
Column 6 gives the peak velocities with respect to the stellar velocities for each detected maser line. 
The dates of the observations (yymmdd) with the corresponding phase of the optical light curve (maximum light = 0.0, 1.0, 2.0, 3.0) are listed in Column 7. 
The optical phases were calculated from the optical data provided by the American Association of Variable Star Observers (AAVSO). 
The thermal lines of $^{28}$SiO $v$ = 0, $J$ = 1--0, $J$ = 2--1 and the high vibrational masers of $v$ = 3, 4, $J$ = 1--0 were not detected 
together with $v$ = 1, 2, $J$ = 3--2 masers. The silicon isotopic lines of $^{29}$SiO $v$ = 0, 1, $J$ = 1--0 and $^{30}$SiO $v$ = 0, $J$ = 1--0 were also not detected. 
The H$_{2}$O maser was not detected at any of the epochs.

Figure~\ref{fig:jkasfig2} shows the peak antenna temperature variations of the SiO $v$ = 1, 2, $J$ = 1--0 masers corresponding to the optical light curve of V407 Cyg as observed 
from March 1, 2010 to July 8, 2014. We can confirm the violent intensity of the variations of the SiO masers by the fact that they were not detected  after the outburst. 
We also confirm the rapid increase in the optical brightness around the outburst as compared to the brightness levels of quiescent phases. 
The peak antenna temperatures of the SiO $v$ = 1, 2, $J$ = 1--0 masers in 2012 occurred with a time delay after the optical maximum. 
In addition, the peak antenna temperature of the SiO $v$ = 2, $J$ = 1--0 maser is always stronger than that of the SiO $v$ = 1, $J$ = 1--0 maser. 

\begin{table}[t!]
\caption{Variation of peak antenna temperature ratios of SiO Masers\label{tab:jkastable3}}
\centering
\begin{tabular}{ccc}

\toprule
Date(phase) & \large $\frac{\rm P.T.(v=2,J=1-0)}{\rm P.T.(v=1,J=1-0)}$ & \large $\frac{\rm P.T.(v=1,J=2-1)}{\rm P.T.(v=1,J=1-0)}$ \\

\midrule
100302(0.00) & 1.39     & $\cdots$  \\
111222(0.92) & 12.45\hh & $\cdots$  \\
120507(1.11) & 3.80     & $\cdots$  \\
120603(1.15) & 3.08     & $\cdots$  \\
121030(1.36) & 1.59     & $\cdots$  \\
130113(1.46) & 2.03     & $\cdots$  \\
130513(1.63) & 1.56     & $\cdots$  \\
130917(1.81) & 2.74     & $\cdots$  \\
131125(1.90) & 2.98     & $\cdots$  \\
140213(2.02) & 2.32     & 0.67      \\
140408(2.09) & 3.23     & 0.95      \\
140605(2.17) & 3.00     & $\cdots$  \\

\bottomrule
\end{tabular}
\end{table}

Table~\ref{tab:jkastable3} gives the values of the peak antenna temperature ratios of the SiO $v$ = 1, 2, $J$ = 1--0 masers. 
In particular, the intensities of the $v$ = 2, $J$ = 1--0 maser are always stronger than those of the $v$ = 1, $J$ = 1--0 masers around the optical maximum 
as shown in Figure~\ref{fig:jkasfig2}. 
The peak antenna temperature ratio of the SiO $v$ = 1 with respect to the $v$ = 2 maser showed the highest value of 12.45 on Dec. 22, 2011 ($\phi$ = 0.92).

\section{Discussion\label{sec:disc}}

\indent\indent As shown in Figure~\ref{fig:jkasfig2} and Table~\ref{tab:jkastable2}, within our three rms detection limits from $\sim$ 0.78 Jy to $\sim$ 1.18 Jy, 
we could not detect the SiO $v$ = 1 and $v$ = 2, $J$ = 1--0 masers during our monitoring interval from April 2, 2010 to June 5, 2010 
after the nova outburst on March 10, 2010. It is clear that the nova outburst causes these non-detections of SiO masers from V407 Cyg. 
Our monitoring results of V407 Cyg can be compared with those of \citet{deg11}. 
\citet{deg11} undertook observations immediately after the outburst from March 16.0, 2010 to May 25.9, 2010 using the NRO 45 m telescope. 
However, they detected weak intensities of SiO $v$ = 1 and $v$ = 2 masers ($\sim$ 0.44--0.59 Jy) and rapid variations of SiO maser line profiles compared to those before the outburst on April 18, 2002. 
The weak maser intensities of $\sim$ 0.44--0.59 Jy were not detected during our KVN observations due to the higher detection limits of three rms levels ($\sim$ 0.78--1.18 Jy). 
\citet{deg11}, with adopting the separation of 2 x 10$^{14}$ cm s$^{-1}$ between the Mira and the white dwarf, explained that the whole SiO maser regions were wiped out by a nova shock on a time scale of $\sim$ 2 weeks after the outburst and the majority of SiO molecules in the optically thin regions around the Mira were dissociated from UV radiation from the shock. 
They also suggested that a part of the SiO maser region was replenished on a time scale of 30 days after the outburst by cool molecular gases expelled by the Mira pulsation. 
Therefore, we can surmise that the SiO maser regions were already wiped out by a nova shock and were not replenished enough for detections with the KVN during our monitoring interval from April 2, 2010 to June 5, 2010. 


\begin{figure*}[t!]
\centering
\includegraphics[width=140mm]{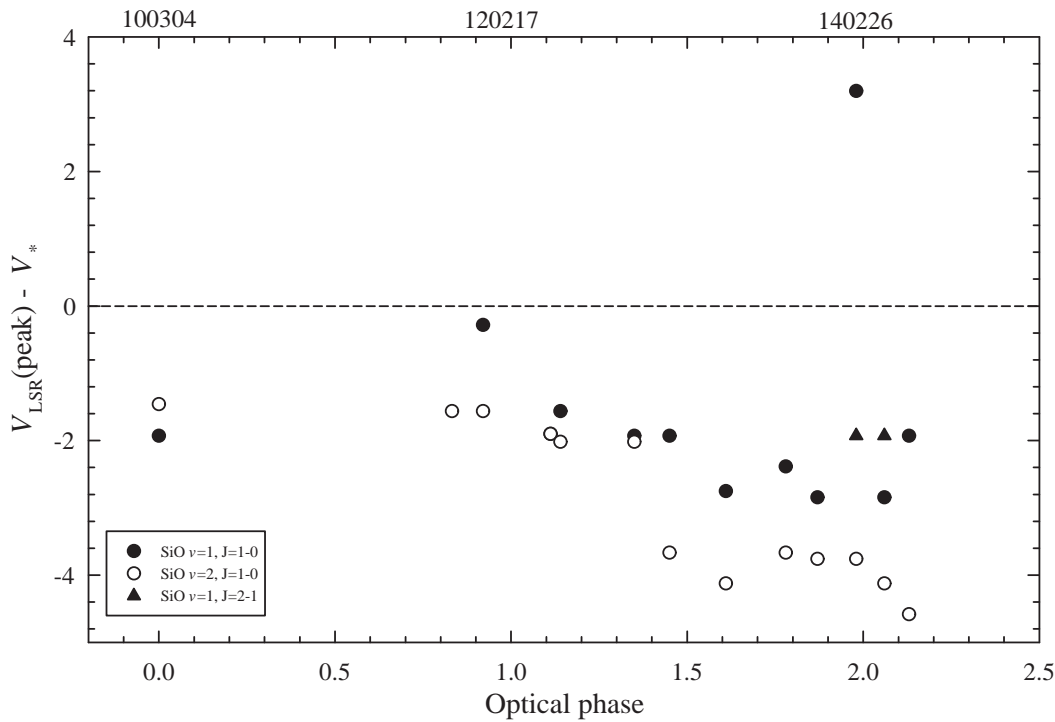}
\caption{Shift of the peak velocities of the detected SiO $v$ = 1 (black circle) and $v$ = 2 (open circle), $J$ = 1--0 and $v$ = 1, $J$ = 2--1 (triangle) masers 
according to the optical phases and observation dates. The Y-axis indicates the peak velocity shift with respect to the stellar velocity $V_{\ast}$ = $-$26.0 km s$^{-1}$. 
\label{fig:jkasfig3}}
\end{figure*}


After the maintenance season of July and August 2010, our monitoring observations were discontinued until September 2011. 
As shown in Figure~\ref{fig:jkasfig2}, the SiO $v$ = 2, $J$ = 1--0 maser was detected from October 20, 2011 ($\phi$ = 0.83). 
Therefore, we could not observe the beginning of the epoch of the SiO maser detections with the KVN single dish after the outburst. 
The spectra after October 20, 2011 showed very different line profiles compared to those of December 2, 2009, three months before the outburst \citep{cho10} 
and March 2, 2010, 8 days before the outburst. The single peak component and two components of the SiO masers with respect to the stellar velocity appeared. 
The peak emission of SiO masers always occurred at blueshifted velocities with respect to the stellar velocity, except for SiO $v$ = 1 on February 13, 2014, 
as shown in the peak velocity variations according to the optical phases (Figure~\ref{fig:jkasfig3}). 
These phenomena may be related to the redistribution of the SiO maser regions after the outburst. 
The quick recovery and high density reservoir of mass outflows through the first Lagrange point (L1) of the binary in the blueshifted part of SiO maser region suggested in 
\citet{deg11} may be associated with these phenomena. 
In Figure~\ref{fig:jkasfig3}, we could not find any tendency of the peak velocity variations associated with stellar pulsation. 
Instead, we found an increasing blueshifted trend during our monitoring interval after the outburst. 
The SiO $v$ = 1, $J$ = 2--1 masers were detected around the maximum optical phases ($\phi$ = 2.02 and 2.09). 
Thus, long-term monitoring observations are required to clarify this trend. 
In addition, two components of the SiO $v$ = 1, 2, $J$ = 1--0 masers with respect to the stellar velocity were detected starting on October 30, 2012. 
These two components may have originated from bipolar outflows (W43A; \citealt{ima05}) or from rotation (for example, IK Tau; \citealt{bob05}) of the SiO masers. 
In order to clarify the originality of these two component features, VLBI observations are required. 
The VLBI observations will allow us to investigate the characteristics of the extent, shape, and velocity gradient of SiO maser regions around the Mira 
including the effect of the hot companion white dwarf which may promote non-spherical mass loss 
through angular momentum and/or gravitational influences \citep{mor87}. 

In addition, although the SiO $v$ = 2 maser requires higher excitation temperature than that of the $v$ = 1 maser, 
the SiO $v$ = 2 intensity of V407 Cyg is always much stronger than that of the $v$ = 1 maser, in contrast to the common Mira variables 
as shown in Figure~\ref{fig:jkasfig2} and Table~\ref{tab:jkastable3}. 
The average values of peak intensity ratios of the SiO $v$ = 2 maser to the $v$ = 1 maser are 1.1 \citep{cho12} and 1.16 \citep{kim10} in Mira variables. 
On the other hand, this value for seven post-AGB stars is 2.47 \citep{yoo14}. 
This indicates that the Mira variable of the V407 Cyg system stands at a very late evolutionary stage compared to common Mira variables. 
It may be related to the development of thick and hot dust shell \citep{nak03,ram12} according to late stellar evolutionary phases, 
which can provide direct SiO $v$ = 2 pumping by absorbing the four micron of dust \citep{lan84}. 

Concerning the non-detections of 22 GHz H$_{2}$O maser from the Mira variable, 
we can speculate that the hot companion white dwarf influences the formation of H$_{2}$O molecules. 
The 22 GHz H$_{2}$O maser regions are located above dust layers, which are distant by more than tenfold from the SiO maser regions \citep{ben96}. 
The 22 GHz H$_{2}$O maser was also not detected from the well-studied symbiotic star R Aqr. 

\section{Summary\label{sec:sum}}

\indent\indent Simultaneous monitoring observations of H$_{2}$O and SiO masers toward V407 Cyg were performed from March 2, 2010 (optical phase $\phi$ = 0.0) eight days before the nova outburst on March 10, 2010 to June 5, 2014 ($\phi$ = 2.13). A summary of our results is as follows. 

\begin{enumerate}
   \item The SiO $v$ = 1, 2, $J$ = 1--0 maser lines exhibited values of 0.51 K ($\sim$ 6.70 Jy) and 0.71 K ($\sim$ 9.30 Jy) on March 2, 2010  ($\phi$ = 0.0), only eight days before the nova outburst, while they were not detected on April 2 ($\phi$ = 0.04), May 5 ($\phi$ = 0.09), May 8 ($\phi$ = 0.09), or June 5 of 2010 ($\phi$ = 0.13) within upper limits from $\sim$0.78 Jy to $\sim$1.18 Jy after the outburst. 
   After the discontinued period of monitoring observations from July of 2010 to September of 2011, we detected the SiO $v$ = 2, $J$ = 1--0 masers starting on October 20, 2011 ($\phi$ = 0.83) and detected the SiO $v$ = 1, $J$ = 1--0 masers starting on December 22, 2011 ($\phi$ = 0.92). 
   These results provide clear evidence of the interaction of the shock from the outburst with the SiO maser regions of the Mira envelope 
   as shown by \citet{abd10} and \citet{deg11}. 
   
   \item The peak emission of the SiO $v$ = 1, 2, $J$ = 1--0 masers always occurred at blueshifted velocities with respect to the stellar velocity 
   except for that of SiO $v$ = 1 on February 13, 2014. These phenomena may be related to the redistribution of the SiO maser regions after the outburst. 
   The quick recovery and high density reservoir of mass outflows through the first Lagrange point (L1) of the binary 
   in the blueshifted part of the SiO maser region suggested in the model by \citet{deg11} may be associated with these phenomena. 
   The peak velocity variations of SiO masers related to the stellar pulsation phases show 
   an increasing blueshifted trend during our monitoring interval after the outburst. 
   
   \item The SiO $v$ = 2 intensity of V407 Cyg is always much stronger than that of the $v$ = 1 maser in contrast to the common Mira variables. 
   This result leads us to conclude that the Mira variable of the V407 Cyg system stands at a very late evolutionary stage compared with common Mira variables.
   
   \item Toward V407 Cyg, we detected the SiO $v$ = 1, $J$ = 2--1 maser lines around the maximum optical phases, February 13, 2014 ($\phi$ = 2.02) 
   and April 6, 2014 ($\phi$ = 2.09), for the first time. 
   However, we could not detect the H$_{2}$O masers at any of the epochs including SiO thermal lines, high transitions of ro-vibrational states, or related isotopic lines. 
\end{enumerate}

\acknowledgments
\indent\indent This work was supported by the Basic and Fusion Research Programs (2011--2015) and also partially supported by the KASI--Yonsei Joint Research Program ``Degree and Research Center Program'' funded by the National Research Council of Science and Technology (NST). 
We are grateful to all staff members in KVN who helped to operate the array, single dish, and to correlate the data. The KVN is a facility operated by KASI (Korea Astronomy and Space Science Institute). The KVN operations are supported by KREONET (Korea Research Environment Open NETwork) which is managed and operated by KISTI (Korea Institute of Science and Technology Information).
In this research, we used information from the AAVSO International Database operated at the AAVSO Headquarters, 25 Birch Street, Cambridge, MA 02138, USA.

\end{document}